\documentclass[sigconf]{acmart}

\usepackage{float}
\usepackage{microtype}
\usepackage{graphicx} 
\usepackage{makecell}
\usepackage{amsmath}

\usepackage{multirow}
\usepackage{subcaption}

\usepackage{algorithm}
\usepackage{algorithmic}

\AtBeginDocument{%
  \providecommand\BibTeX{{%
    \normalfont B\kern-0.5em{\scshape i\kern-0.25em b}\kern-0.8em\TeX}}}

\copyrightyear{2023}
\acmYear{2023}
\setcopyright{acmlicensed}\acmConference[KDD '23]{Proceedings of the 29th ACM SIGKDD Conference on Knowledge Discovery and Data Mining}{August 6--10, 2023}{Long Beach, CA, USA}
\acmBooktitle{Proceedings of the 29th ACM SIGKDD Conference on Knowledge Discovery and Data Mining (KDD '23), August 6--10, 2023, Long Beach, CA, USA}
\acmPrice{15.00}
\acmDOI{10.1145/3580305.3599923}
\acmISBN{979-8-4007-0103-0/23/08}

\begin{document}

\title{UA-FedRec: Untargeted Attack on Federated News Recommendation}

\author{Jingwei Yi}
\affiliation{
    \institution{University of Science\\and Technology of China}
    \country{}
}
\email{yjw1029@mail.ustc.edu.cn}

\author{Fangzhao Wu}
\affiliation{
    \institution{Microsoft Research Asia}
    \country{}
}
\email{fangzwu@microsoft.com}

\author{Bin Zhu}
\affiliation{
    \institution{Microsoft Research Asia}
    \country{}
}
\email{binzhu@microsoft.com}

\author{Jing Yao}
\affiliation{
    \institution{Microsoft Research Asia}
    \country{}
}
\email{jingyao@microsoft.com}

\author{Zhulin Tao}
\affiliation{
    \institution{Communication University of China}
    \country{}
}
\email{taozhulin@gmail.com}

\author{Guangzhong Sun}
\affiliation{
    \institution{University of Science\\and Technology of China}
    \country{}
}
\email{gzsun@ustc.edu.cn}

\author{Xing Xie}
\affiliation{
    \institution{Microsoft Research Asia}
    \country{}
}
\email{xingx@microsoft.com}


\renewcommand{\shortauthors}{Yi, et al.}

\begin{abstract}

News recommendation is essential for personalized news distribution.
Federated news recommendation, which enables collaborative model learning from multiple clients without sharing their raw data, is a promising approach for preserving users' privacy.
However, the security of federated news recommendation is still unclear.
In this paper, we study this problem by proposing an untargeted attack on federated news recommendation called UA-FedRec.
By exploiting the prior knowledge of news recommendation and federated learning, UA-FedRec can effectively degrade the model performance with a small percentage of malicious clients.
First, the effectiveness of news recommendation highly depends on user modeling and news modeling.
We design a news similarity perturbation method to make representations of similar news farther and those of dissimilar news closer to interrupt news modeling, and propose a user model perturbation method to make malicious user updates in opposite directions of benign updates to interrupt user modeling.  
Second, updates from different clients are typically aggregated with a weighted average based on their sample sizes. We propose a quantity perturbation method to enlarge sample sizes of malicious clients in a reasonable range to amplify the impact of malicious updates. 
Extensive experiments on two real-world datasets show that UA-FedRec can effectively degrade the accuracy of existing federated news recommendation methods, even when defense is applied.
Our study reveals a critical security issue in existing federated news recommendation systems and calls for research efforts to address the issue.
Our code is available at \url{https://github.com/yjw1029/UA-FedRec}.
\end{abstract}

\ccsdesc[500]{Information systems~Collaborative filtering}
\ccsdesc[500]{Security and privacy~Privacy protections}
\keywords{Untargeted attack, Federated learning, News recommendation}



\maketitle

\section{Introduction}
Nowadays, a large amount of news is posted on the Web every day, leading to severe information overload.
Personalized news recommendation is proposed. It aims to recommend news according to user interests~\cite{okura2017embedding,ijcai2019-536,wang-etal-2020-fine,qi-etal-2021-pp,wu-etal-2019-neural-news,an-etal-2019-neural}.
Most personalized news-recommendation approaches have three components: a news model, a user model, and a click prediction module.
The news model learns news representations from news textual information.
The user model learns user representations from users' historical clicked news.
The click prediction module predicts click scores for each user-and-news-representation pair.
However, most news recommendation methods rely on centralized storage, which raises concerns about user privacy.
Moreover, some privacy laws, such as GDPR\footnote{https://gdpr-info.eu/} and CCPA\footnote{https://oag.ca.gov/privacy/ccpa}, are enacted to protect user privacy.
It may not be able to train models with centralized user data in the future.

Federated learning (FL) is a technology that enables multiple clients to collaboratively train a model without sharing their training data~\cite{mcmahan2017communication}.
Several federated news-recommendation methods are proposed for privacy-preserving news recommendation~\cite{qi-etal-2020-privacy, yi-etal-2021-efficient, qi-etal-2021-uni-fedrec}. 
~\citet{qi-etal-2020-privacy} propose a privacy-preserving news recommendation method, called FedRec, based on federated learning.
In FedRec, a central server keeps a global news recommendation model and distributes it to a group of randomly sampled clients in each round.
Selected clients train their local models and upload model updates to the server. The server updates the global news recommendation model by aggregating received model updates.
~\citet{yi-etal-2021-efficient} propose Efficient-FedRec, an efficient federated learning framework for privacy-preserving news recommendation.
In Efficient-FedRec, the news recommendation model is decomposed into a large news model maintained in the server and a light-weight user model shared among the server and clients, while news representations and the user model are communicated between the server and clients.
~\citet{qi-etal-2021-uni-fedrec} propose a unified news recommendation framework. It contains recall and ranking stages, and can train models and serve users in a privacy-preserving way.

Although these federated news recommendation methods can protect user privacy, the security of federated news recommendation systems is unclear.
Since clients need to submit model updates to the central server in federated news recommendation systems, it is possible that an attacker controls multiple malicious clients to submit poisoned updates to attack the global news recommendation model, resulting in degraded performance or preventing convergence of the global news recommendation model. Such attacks are known as untargeted attacks.
An untargeted attack on federated news recommendation can impact a large number of benign clients/users and severely deteriorate the user experience.
Therefore, it is necessary to study potential attacks on and effective defenses for federated news recommendation systems.
In this paper, we propose an untargeted attack, called \emph{UA-FedRec}, on federated news recommendation systems.
By fully exploiting the prior knowledge of news recommendation and federated learning, UA-FedRec can efficiently degrade the global model performance with a small percentage of malicious clients.
Since the performance of news recommendation models highly depends on the accuracy of user modeling and news modeling~\cite{10.1145/3097983.3098108,wu2019neural,an-etal-2019-neural,10.1145/3292500.3330665},
we design a news similarity perturbation method to make representations of similar news farther and those of dissimilar news closer and propose a user model perturbation method to make malicious updates neutralize benign updates.
Additionally, since updates from different clients are aggregated in vanilla federated learning with weighted-averaging based on their sample sizes,  
we amplify the impact of malicious updates by proposing a quantity perturbation method that enlarges sample sizes of malicious clients in a reasonable range.
Extensive experiments on two real-world datasets prove UA-FedRec's effectiveness, even under defenses. Our study reveals a critical security issue in existing federated news recommendation systems, which should draw the attention of researchers in the field.
The main contributions of this paper can be summarized as follows:
\begin{itemize}
    \item We present the first study on untargeted attacks against federated news recommendation.
    \item We propose UA-FedRec, an efficient untargeted attack on federated news recommendation systems. It requires a small percentage of malicious clients and is thus more practical.
    \item Extensive experiments on two real-world datasets prove UA-FedRec's effectiveness, even under defenses.
    We raise a critical vulnerability in existing federated news recommendation systems, which should draw the attention of researchers in the field.

\end{itemize}
\section{Related Work}
\subsection{Personalized News Recommendation}
Personalized news recommendation is a critical way to personalize news distribution and alleviate the information overload problem.
Multiple news recommendation methods have been proposed recently~\cite{okura2017embedding,ijcai2019-536,wang-etal-2020-fine,qi-etal-2021-pp,10.1145/3404835.3463069,10.1145/3292500.3330665,qi2021personalized}.
Generally, there are three core components in news recommendation methods: a news model, a user model, and a click prediction module.
The news model is used to learn news representations from news textual information.
For example, ~\citet{10.1145/3178876.3186175} propose to learn news representations with a knowledge-aware convolutional network (KCNN) and a max-pooling layer.
~\citet{wu-etal-2019-neural-news} use the combination of multi-head self-attention and additive attention to learn news representations.
~\citet{10.1145/3404835.3463069} apply pre-trained language model in the news model to empower its semantic understanding ability.
The user model is used to learn user representations from users' historical clicked news representations.
For example, ~\citet{10.1145/3292500.3330665} apply user embeddings as the query of an additive attention layer to learn user representations.
~\citet{an-etal-2019-neural} use a GRU network to capture short-term user interests, and use user embeddings to capture long-term user interests.
~\citet{qi2021personalized} apply a candidate-aware additive attention network to learn user representations.
Click prediction model computes a click score given a pair of user and candidate news representation, which can be implemented by dot product~\cite{10.1145/3404835.3463069}, cosine similarity~\cite{kumar2017deep}, or MLP network~\cite{10.1145/3178876.3186175}.

\subsection{Federated Recommendation System}
Federated learning is a technique in that multiple clients collaboratively train a global model without sharing their private data~\cite{mcmahan2017communication}.
It performs the following three steps in each round.
First, the central server distributes the current global model to a group of randomly sampled clients.
Second, each selected client trains the local model with local private data and
sends the model update and the number of training samples to the central server.
Third, the server aggregates the model updates received from clients to update the global model according to a specific aggregation rule. 
In FedAvg~\cite{mcmahan2017communication}, updates are weightedly averaged based on sample sizes of clients.

Federated learning has been applied to build privacy-preserving recommendation systems~\cite{Liang_Pan_Ming_2021,10.1145/3397271.3401081,10.1145/3394486.3403176,9170754,9174297,10.1145/3383313.3411528}.
~\citet{Ammaduddin2019FederatedCF} propose federated collaborative filtering (FCF).
In FCF, clients use their local private data to compute updates of local user embeddings and item embeddings in the CF model, and submit updates of item embeddings.
~\citet{shin2018privacy} propose secure federated matrix factorization (FMF).
FMF is similar to FCF, but clients compute local updates according to the matrix factorization algorithm.
~\citet{qi-etal-2020-privacy} propose FedRec, a privacy-preserving method for news recommendation model training.
In FedRec, clients utilize their local data to compute local updates of the news recommendation model and upload the updates to the central server.
The central server further aggregates the updates to update the global model.

\subsection{Poisoning Attacks}
Poisoning attacks 
interfere with model training via manipulating
input samples or model parameters to achieve a certain malicious goal. They can be divided into three categories according to the goal to achieve: targeted attacks, backdoor attacks, and untargeted attacks.
Targeted attacks~\cite{bhagoji2019analyzing} aim to cause misprediction on a specific set of target samples while maintaining the same prediction on the rest of samples.
Backdoor attacks~\cite{liu2017trojaning,bagdasaryan2020backdoor,DBLP:conf/nips/WangSRVASLP20,Xie2020DBA:} aim to cause misprediction only when the backdoor trigger is applied.
Untargeted attacks~\cite{NEURIPS2019_ec1c5914,10.5555/3489212.3489304} aim to degrade the performance on arbitrary input samples.
Poisoning attacks can also be divided into two categories according to the attack method: data poisoning attacks and model poisoning attacks.
Data poisoning attacks~\cite{biggio2011support,259745,8975792,pmlr-v97-mahloujifar19a} manipulate input samples, while
model poisoning attacks~\cite{bhagoji2019analyzing,NEURIPS2019_ec1c5914,10.5555/3489212.3489304} directly manipulate model parameters.

Several data poisoning attack methods have been proposed on recommendation systems~\cite{10.5555/3157096.3157308,10.1145/1278366.1278372,yang2017fake,10.1145/3274694.3274706}.
These attacks involve injecting fake user-item interactions into the training dataset to increase the exposure rate of the target item.
For instance, \citet{10.1145/3274694.3274706} propose an attack on graph-based recommendation systems by formulating it as an optimization problem.
However, these methods assume that the adversary has access to the complete recommendation system history, which may not be feasible in practice.
To address this issue, \citet{10.1145/3447548.3467233} design an attack based on incomplete data.
It is important to note that all the aforementioned attacks are applicable to centralized recommendation systems.
Recently, \citet{zhang2021pipattack} introduce PipAttack, a poisoning attack specifically targeting federated recommendation systems. PipAttack trains a popularity classifier and generates perturbed updates to enhance the popularity of the target item. However, PipAttack is primarily designed for promoting one or more specific target items and is not effective for untargeted attacks. Another approach proposed by \citet{10.1145/3534678.3539119} is FedAttack, which leverages negative sampling of items with representations similar to those of the malicious user to degrade the performance of federated recommendation systems.
Notably, FedAttack is only effective for ID-embedding-based recommendation methods.
To the best of our knowledge, there are currently no effective untargeted attacks specifically designed for the federated news recommendation scenario.


Recently, several untargeted attacks on federated learning have been proposed~\cite{NEURIPS2019_ec1c5914,10.5555/3489212.3489304}.
Label flipping~\cite{10.5555/3489212.3489304} is an untargeted data poisoning attack on federated learning by flipping labels of training samples at malicious clients.
Some model poisoning attacks on federated learning have been proposed to directly manipulate model updates, which can usually achieve better performance.
LIE~\cite{NEURIPS2019_ec1c5914} adds a small mount of noise on each dimension of the average of benign updates, with the noise being small enough to circumvent defense methods.
~\citet{10.5555/3489212.3489304} propose to add noise in the opposite direction from the average of benign updates.
Besides, they tailor the attack algorithm to evade defenses. 
However, these untargeted attacks are usually based on a large percentage of malicious clients, which is not practical for federated recommendation systems.
\begin{figure*}[!t]
  \centering
  \includegraphics[width=0.99\textwidth]{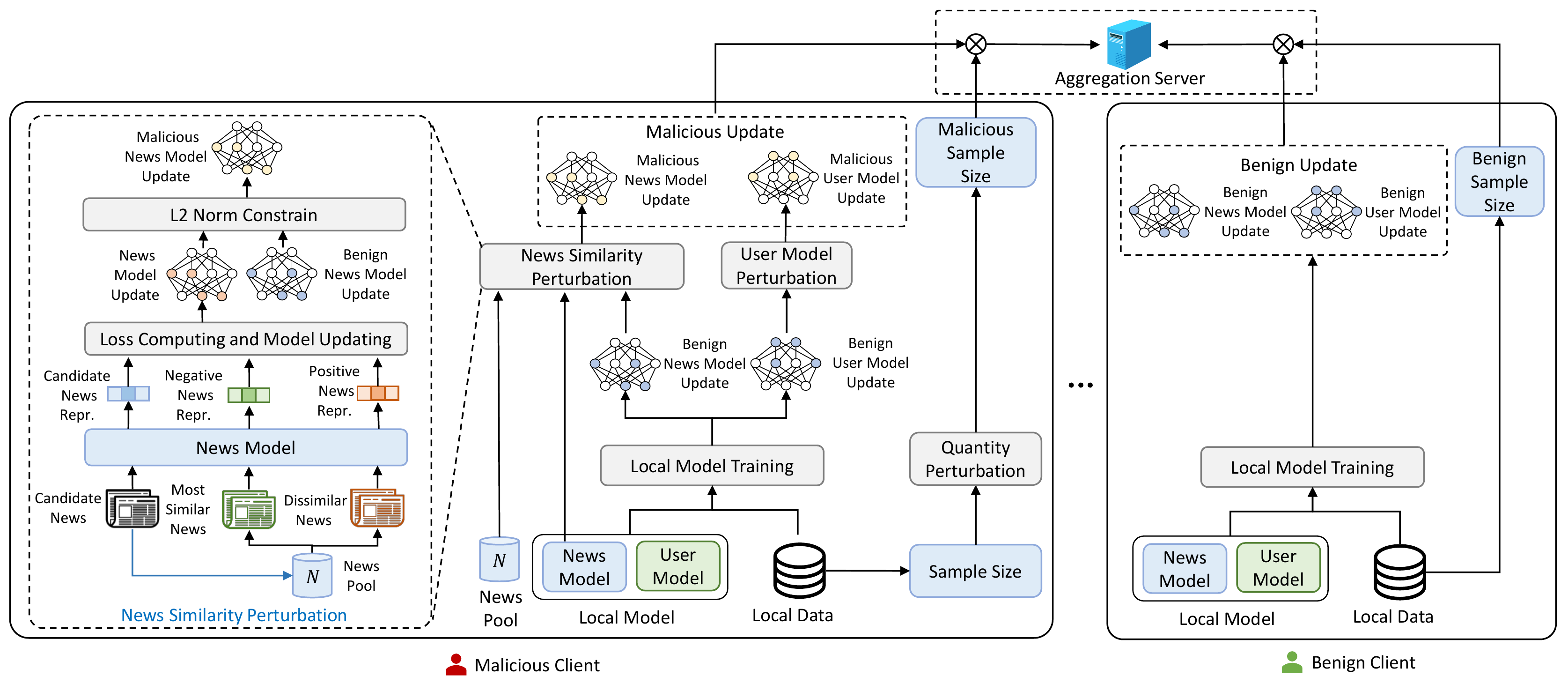}
  \caption{The framework of our UA-FedRec method.}
  \label{fig:uattack4rec}
\end{figure*}

\subsection{Problem Formulation}
\label{sec:problem-formulation}
Denote the news set as $\mathcal{N}=\{n_1, n_2, ... n_L\}$, where $L$ is the number of pieces of news.
Each piece of news $n_i$ is presented by its title $t_i$.
Denote $\mathcal{U}=\{u_1, u_2, ... u_N\}$ as the total clients participating in federated model training, where $N$ is the number of clients.
Given a user $u_j$, his private click data $\mathcal{B}_j$ is stored in his local device.
In federated news recommendation, these $N$ clients collaboratively train a global news recommendation model $\boldsymbol \Theta$.
In each round, the central server randomly selects $k$ clients.
Each selected client trains his local news recommendation model with his local dataset.
The difference between the updated model and the global model received from the server is denoted as the model update $\textbf{g}$. Model updates are uploaded by the selected clients and further aggregated by the central server.
Among the $N$ clients, we assume there are $m$ malicious clients controlled by an attacker.
The malicious clients are denoted as $\mathcal{U}_m = \{u_1, u_2, ... u_m\} \subseteq \mathcal{U}$.
The attacker aims to degrade the resulting global model's performance on any input samples by uploading malicious model updates $\textbf{g}^m$ from selected malicious clients.

\subsection{Threat Model}
\noindent \textbf{Attacker's Objective.}
The attacker's objective is to degrade the performance of the federated news recommendation system on arbitrary input samples, i.e., it is an untargeted attack on a federated news recommendation system.

\noindent \textbf{Attacker's Capability.}
As mentioned in Section~\ref{sec:problem-formulation}, there are $m$ malicious clients, controlled by an attacker, among $N$ clients participating in model training.
Since a recommendation system generally has millions of users in practice, we believe that a reasonable percentage of malicious clients should be up to 1\%.
The attacker can manipulate model updates of malicious clients to degrade the performance of the global model.

\noindent \textbf{Attacker's Knowledge.}
We assume that the attacker has full access to the code, local model, and benign datasets on devices of malicious clients.
Additionally, we assume the attacker has the information of some pieces of news, such as news titles.
Since clients in federated news recommendation do not share their local data, we assume that the attacker has only partial knowledge of the data distribution.
Since the server might not release its aggregation code, we assume the attacker does not know the aggregation rule used by the server.
Meanwhile, we assume malicious clients can communicate and collaborate to attack the global recommendation model.

\subsection{Basic News Recommendation Model}
\label{sec:base-news}
FedRec~\cite{qi-etal-2020-privacy} is compatible with the majority of news recommendation models.
For generality, our UA-FedRec is agnostic of the news recommendation model structure.
A news recommendation model is generally composed of three core components: a news model, a user model, and a click prediction model.
Given a piece of news $n$, the news model generates the news representation $\textbf{n}$ from the news title.
We will conduct experiments on two models, NRMS~\cite{wu-etal-2019-neural-news} and LSTUR~\cite{an-etal-2019-neural}.
In NRMS, the news model is implemented with the combination of a multi-head self-attention network and an additive attention layer.
In LSTUR, the news model is composed of a convolutional network and an additive attention layer.
Given historical news representations $[\textbf{n}_1, \textbf{n}_2 ... \textbf{n}_s]$ of a user $u$, the user encoder learns the user representation $\textbf{u}$.
NRMS applies the combination of a user-level multi-head self-attention network and an additive attention network to learn user representations.
LSTUR uses user-ID embeddings to capture users' short-term interests and uses a GRU network to capture users' long-term interests.
The click prediction model computes click score $c$ for each pair of user and candidate news representation, which is implemented by dot product in both NRMS and LSTUR.

Both NRMS~\cite{wu-etal-2019-neural-news} and LSTUR~\cite{an-etal-2019-neural} apply a negative sampling strategy to compute loss.
For each clicked piece of news, $P$ unclicked pieces of news are sampled in the same impression. For the $P+1$ samples, we denote their click scores 
as $\{c_1, c_2, ..., c_{P+1}\}$ and their click labels 
as $\{y_1, y_2, ..., y_{P+1}\}$.
The click probability of the $i$-th piece of news is computed as $p_i = exp(c_i) / \sum_{j=1}^{P+1}exp(c_j))$, and the loss of this sample is computed as $\mathcal{L} = -\sum_{i=1}^{P+1}y_i \times log(p_i)$.
For a benign client $u_j$, the average loss over all samples in his local dataset is computed, which is defined as $\mathcal{L}_j = \sum_{i \in \mathcal{B}_j} \mathcal{L}_j^i /\|\mathcal{B}_j\|$, where $\mathcal{L}_j^i$ is the loss of the sample $i$.
Loss $\mathcal{L}_j$ is used to compute an update from client $u_j$, which is denoted as $\textbf{g}_j$.

\subsection{Framework of UA-FedRec}
In this subsection, we introduce our UA-FedRec on federated news recommendation.
The overall framework is shown in Figure~\ref{fig:uattack4rec}.
It is composed of three core components: user model perturbation, news similarity perturbation, and quantity perturbation.
Their details are described as follows.

\subsubsection{User model perturbation}
The user model perturbation is used to generate malicious updates for the user model.
In UA-FedRec, we leverage the prior knowledge in news recommendation that the performance of news recommendation highly depends on user modeling and perturb updates of the user model in opposite directions of benign updates.
First, we estimate benign updates from benign datasets in the devices of malicious clients.
Specifically, for each malicious client $u_i \in \mathcal{U}_m$, we compute a benign update following the steps described in Section~\ref{sec:base-news}. 
The benign user model update of client $u_i$ is denoted as $\textbf{g}_i^u$.
Then we average the benign user model updates of all malicious clients to estimate a benign user model update: $\boldsymbol\mu_u = \sum_{1\leq i\leq m} \textbf{g}_i^u/m$.
Second, we compute the direction of the estimated benign user model update, $\textbf{s}_u = sign(\boldsymbol\mu_u)$.
We also compute the standard deviation of the benign user model updates of all malicious clients, which is denoted as $\boldsymbol\sigma_u$.
To circumvent defenses, a malicious user update should not be too far away from a benign update. To meet this requirement, the malicious update from a malicious client is computed as $\textbf{g}_u^m = \boldsymbol \mu_u - \lambda_1\, \textbf{s}_u\odot\boldsymbol\sigma_u$, where $\lambda_1 \leq 4$ and $\odot$ stands for the element-wise product operation.

\subsubsection{News similarity perturbation}
The news similarity perturbation is used to generate malicious updates for the news model.
It is motivated by the prior knowledge of news recommendation that news modeling is critical for news recommendation.
For example, a user who read ``Best PS5 games: top PlayStation 5 titles to look forward to'' likely also read ``Marvel's Avengers game release date, news, trailers and first impressions'', but is less likely to click ``The Cost of Trump's Aid Freeze in the Trenches of Ukraine's War''.
For a good news recommendation model, the second news's representation should be close to the first news's representation in the vector space, while the third news's representation should be far away from the first news's representation in the vector space.
Therefore, we design our news similarity perturbation to make representations of similar news farther and those of dissimilar news closer.

First, we inference news representations and search the closest and farthest pieces of news for each piece of news.
For the $i$-th piece of news $n_i$, its closest and farthest pieces of news, denoted as $n_i^n$ and $n_i^f$, respectively, can be computed as follows:
\begin{equation}
    \begin{aligned}
        n_i^f = \min_{n_j\in \mathcal{N}, j\neq i}{\textbf{n}_i^T \textbf{n}_j}, \
        n_i^n = \max_{n_j\in \mathcal{N}, j\neq i}{\textbf{n}_i^T \textbf{n}_j},
    \end{aligned}
    \label{eq::news_neighbor}
\end{equation}
where $\textbf{n}_i$ and $\textbf{n}_j$ are news representations for the $i$-th and the $j$-th pieces of news, respectively. 
Computing all news representations and selecting neighbors in each round are time-consuming.
To reduce complexity, we assume that distances between news representations do not change significantly in $K$ rounds, and thus update the selected news neighbors once every $K$ rounds.
We empirically validate the assumption in Appendix~\ref{sec:distances-no-change}.

Second, we enlarge the MSE loss between $\textbf{n}_i$ and $\textbf{n}_i^n$ and reduce the MSE loss between $\textbf{n}_i$ and $\textbf{n}_i^f$. 
The news similarity perturbation loss is computed as follows:
\begin{equation}
    \mathcal{L}_n=\sum_{n_i\in \mathcal{N}}(\textbf{n}_i - \textbf{n}_i^f)^T(\textbf{n}_i - \textbf{n}_i^f)-(\textbf{n}_i - \textbf{n}_i^n)^T(\textbf{n}_i - \textbf{n}_i^n).
\label{eq::news_loss}
\end{equation}
The local model is updated using the loss in Eq.~\ref{eq::news_loss} with the backpropagation algorithm to get news model update $\textbf{g}_n$. 
To evade detection, we constrain the $L_2$ norm of a malicious news model update not too far away from the $L_2$ norm of benign news model updates. We estimate benign updates in the following way.
For each malicious client $u_i \in \mathcal{U}_m$, we compute its benign news model update $\textbf{g}_i^n$ using its local benign dataset according to the steps described in Section~\ref{sec:base-news}.
We then compute the average and the standard deviation of the $L_2$ norm of benign updates from all malicious clients, denoted as $\mu_n$ and $\sigma_n$, respectively.
Assuming the $L_2$ norm of benign updates follow the Gaussian distribution, we set a reasonable maximum $L_2$ norm of malicious news model updates as $\mu_n + \lambda_2\sigma_n$.
The final malicious news model update is thus computed as:
\begin{equation}
    \textbf{g}_n^m=\frac{\textbf{g}_n}{max(1, ||\textbf{g}_n||_2/(\mu_n + \lambda_2\sigma_n))}.
\end{equation}

\subsubsection{Quantity perturbation}
In most federated learning methods, updates from different clients are aggregated with a weighted average based on their sample sizes. To exploit this prior knowledge, we enlarge sample sizes of malicious clients in sending to the server 
to magnify the impact of malicious updates. Generated malicious sample sizes should be sufficiently large to enhance the influence of malicious updates, but should also be small enough to evade detection.
Unlike some other federated learning scenarios, sample sizes vary significantly across clients in the recommendation scenario~\cite{DBLP:conf/ijcai/XuYTTL15,10.1145/3269206.3271710}.
We leverage this characteristic to enlarge sample sizes of malicious clients in the following way.
Denote benign sample sizes of malicious clients as $\{s_1, s_2, ... s_m\}$.
We compute their average and standard deviation, denoted as $\mu_s$ and $\sigma_s$, respectively.
The final sample size submitted to the central server by a malicious client is $\lfloor \mu_s+\lambda_3\sigma_s \rfloor$ , where $0 \leq \lambda_3 \leq 4$. 
The mean, standard deviation, and maximum value of benign sample sizes are 3.14, 3.00, and 62, respectively, for the MIND dataset, and are 23.35, 45.24, and 691, respectively, for the Feeds dataset.
These observations demonstrate selecting $\lambda_3 \leq 4$ is reasonable.

\subsection{Complexity Analysis}

\begin{table}[!t]
\centering
\caption{Complexity analysis of different methods, where $m$ is the number of malicious clients, $q$ is the possibility that there is at least 1 malicious client being sampled, $p$ is the average number of malicious clients being sampled in each round, $D$ is the average local data size of clients, $\boldsymbol \Theta$ is the news recommendation model, $\boldsymbol \Theta_n$ is the news model, $\boldsymbol \Theta_u$ is the user model, $K$ is the number of rounds to update the selected news neighbors, $d$ is the dimension of news and user representations, $L$ is the number of news.
}
\scalebox{0.95}{
\begin{tabular}{lc}
\Xhline{1.5pt}
Method & Complexity \\ \hline
LF & $O(pD\|\boldsymbol \Theta\|)$ \\ 
Pop  &$O(pD\|\boldsymbol \Theta\|)$ \\
FedAttack  & $O(pD\|\boldsymbol \Theta\|) + qL\|\boldsymbol \Theta_n\| + p\|\boldsymbol \Theta_u\| + pLd)$ \\
Gaussian  & $O(qmD\|\boldsymbol \Theta\|)$ \\
LIE  & $O(qmD\|\boldsymbol \Theta\|)$  \\
FANG & $O(qmD\|\boldsymbol \Theta\|)$  \\
UA-FedRec & $O(qmD\|\boldsymbol \Theta\| + qL\|\boldsymbol \Theta_n\| + qL^2d/K)$ \\
\Xhline{1.5pt}
\end{tabular}
}
\label{tab:complexity}
\end{table}

In this section, we analyze the computational complexity of UA-FedRec.
As shown in Table~\ref{tab:complexity}, the complexity of UA-FedRec consists of three components.
The first component ($qmD\|\boldsymbol \Theta\|$) represents the cost associated with estimating benign updates using local benign data from the malicious user.
The second component ($qL\|\boldsymbol \Theta_n\|$) indicates the cost of updating the news model based on Eq.~\ref{eq::news_loss}.
The third component ($qL^2d/K$) signifies the cost of updating the nearest and farthest news for each news item every K rounds based on Eq.~\ref{eq::news_neighbor}.


\begin{table}[!t]
\centering
\caption{Detailed statistics of MIND and Feeds.}
\scalebox{0.98}{
\begin{tabular}{c|cc}
\Xhline{1.5pt}
                     & MIND          & Feeds          \\ \hline
\#news               & 65,238              & 643,177        \\
\#users              & 94,057              & 10,000         \\
\#impressions        & 230,117             & 320,578        \\
\#positive samples & 347,727             & 437,072        \\
\#negative samples & 8,236,715           & 6,621,187      \\
\Xhline{1.5pt}
\end{tabular}
}
\label{tab:stat}
\end{table}

\begin{table*}[!t]
\centering
\caption{Attack performance of different attack methods with no defense.}
\scalebox{0.95}{
\begin{tabular}{c|c|cccc|cccc}
\Xhline{1.0pt}
        Base                & \multirow{2}{*}{Methods} & \multicolumn{4}{c|}{MIND}                         & \multicolumn{4}{c}{Feeds}    \\ \cline{3-10} 
        Model               &                          & AUC        & MRR        & nDCG@5     & nDCG@10    & AUC & MRR & nDCG@5 & nDCG@10 \\ \hline
\multirow{6}{*}{NRMS}       & No Attack         & 66.73±0.13 & 32.34±0.15 & 35.05±0.14 & 40.75±0.12 & 65.05±0.09 & 31.92±0.10 & 34.39±0.12 & 42.15±0.10 \\
                            & LF~\cite{10.5555/3489212.3489304}           & 66.69±0.15 & 32.26±0.10 & 34.97±0.10 & 40.69±0.09 & 64.90±0.11 & 31.78±0.10 & 34.20±0.13 & 42.00±0.13 \\
                            & Pop~\cite{zhang2021pipattack}          & 66.72±0.23 & 32.34±0.12 & 35.05±0.12 & 40.74±0.12 & 64.99±0.18 & 31.87±0.13 & 34.33±0.16 & 32.11±0.16 \\
                            & FedAttack~\cite{10.1145/3534678.3539119} & 66.65±0.22 & 32.27±0.13 & 34.98±0.16 & 40.68±0.14 & 65.04±0.13 & 31.82±0.14 & 34.29±0.15 & 42.08±0.15 \\
                            & Gaussian~\cite{10.5555/3489212.3489304}     & 66.64±0.17 & 32.33±0.13 & 35.02±0.15 & 40.71±0.12 & 64.87±0.17 & 31.82±0.11 & 34.27±0.15 &       42.04±0.12 \\
                            & LIE~\cite{NEURIPS2019_ec1c5914}          & 59.52±0.43 & 27.69±0.26 & 29.43±0.27 & 35.03±0.27 & 61.63±0.25 & 29.19±0.15 & 30.85±0.19 & 38.85±0.21 \\
                            & Fang~\cite{10.5555/3489212.3489304}         & 62.92±0.71 & 29.64±0.48 & 31.83±0.57 & 37.52±0.58 & 61.04±0.26 & 28.74±0.16 & 30.33±0.19 & 38.31±0.19 \\ \cline{2-10} 
                            & UA-FedRec      & \textbf{55.81}±0.34 & \textbf{25.08}±0.37 & \textbf{26.19}±0.37 & \textbf{31.79}±0.35 & \textbf{58.96}±0.61 & \textbf{27.13}±0.52 & \textbf{28.30}±0.63 & \textbf{36.39}±0.58 \\ \hline
\multirow{6}{*}{LSTUR}      & No Attack         & 66.67±0.09 & 32.30±0.12 & 34.97±0.11 & 40.67±0.11 & 65.17±0.04 & 31.91±0.08 & 34.39±0.13 & 42.19±0.08 \\
                            & LF~\cite{10.5555/3489212.3489304}           & 66.63±0.09 & 32.24±0.08 & 34.87±0.10 & 40.58±0.10 & 65.12±0.13 & 31.80±0.14 & 34.27±0.17 & 42.09±0.15        \\
                            & Pop~\cite{zhang2021pipattack}          & 66.81±0.14 & 32.40±0.11 & 35.07±0.14 & 40.76±0.13 & 65.30±0.05 & 32.01±0.04 & 34.50±0.05 & 42.32±0.04        \\
                            & FedAttack~\cite{10.1145/3534678.3539119} & 66.87±0.06 & 32.40±0.02 & 35.03±0.03 & 40.75±0.02 & 65.10±0.03 & 31.89±0.03 & 34.35±0.05 & 42.17±0.04 \\ 
                            & Gaussian~\cite{10.5555/3489212.3489304}     & 66.69±0.14 & 32.26±0.11 & 34.90±0.13 & 40.62±0.13 & 65.15±0.03 & 31.91±0.02 & 34.42±0.03 & 42.18±0.02        \\
                            & LIE~\cite{NEURIPS2019_ec1c5914}          & 63.56±0.20 & 29.99±0.25 & 32.20±0.24 & 37.91±0.23 & 63.93±0.57 & 30.78±0.41 & 32.99±0.54 & 40.91±0.50 \\
                            & Fang~\cite{10.5555/3489212.3489304}         & 63.87±0.35 & 30.33±0.26 & 32.57±0.33 & 38.25±0.30 & 61.81±0.67 & 29.17±0.47 & 30.93±0.57 & 38.92±0.58        \\ \cline{2-10} 
                            & UA-FedRec      & \textbf{54.33}±0.69 & \textbf{24.37}±0.70 & \textbf{25.30}±0.66 & \textbf{30.96}±0.56 & \textbf{59.36}±0.39 & \textbf{27.25}±0.31 & \textbf{28.52}±0.36 & \textbf{36.64}±0.35 \\ \Xhline{1.0pt}
\end{tabular}
}
\label{tab:exp-no-defense}
\end{table*}

\section{Experimental Evaluation}

\begin{table*}[!t]
\centering
\caption{Performance of different defense methods without any attack.}
\scalebox{0.92}{
\begin{tabular}{c|c|cccc|cccc}
\Xhline{1.0pt}
                Base        & \multirow{2}{*}{Methods} & \multicolumn{4}{c|}{MIND}                         & \multicolumn{4}{c}{Feeds}    \\ \cline{3-10} 
                Model       &                          & AUC        & MRR        & nDCG@5     & nDCG@10    & AUC & MRR & nDCG@5 & nDCG@10 \\ \hline
\multirow{6}{*}{NRMS}       & No Defense               & 66.73±0.13 & 32.34±0.15 & 35.05±0.14 & 40.75±0.12 & 65.05±0.09 & 31.92±0.10 &                                    34.39±0.12 & 42.15±0.10 \\
                            & Median~\cite{yin2018byzantine}                   & 56.05±0.24 & 25.45±0.07 & 26.50±0.07 & 32.06±0.10 & 60.56±0.12 & 28.32±0.18 &      29.82±0.22 & 37.85±0.17 \\
                            & Trimmed-Mean~\cite{yin2018byzantine}             & 63.64±0.25 & 30.00±0.22 & 32.12±0.24 & 37.85±0.23 & 61.31±0.25 & 28.84±0.18 & 30.46±0.23 & 38.52±0.21 \\
                            & Krum~\cite{NIPS2017_f4b9ec30}                    & 56.97±0.03 & 25.84±0.18 & 27.15±0.19 & 32.82±0.12 & 62.15±0.29 & 29.49±0.32 & 31.37±0.37 & 39.35±0.34 \\
                            & Multi-Krum~\cite{NIPS2017_f4b9ec30}             & 65.80±0.17 & 31.66±0.10 & 34.23±0.11 & 39.93±0.12 & 62.51±0.08 & 29.73±0.06 & 31.62±0.07 & 39.63±0.08 \\
                            & Norm-Bounding~\cite{sun2019can}                 & 66.92±0.19 & 32.44±0.13 & 35.18±0.14 & 40.88±0.14 & 64.97±0.05 & 31.84±0.09 & 34.31±0.10 & 42.08±0.09 \\ \hline
\multirow{6}{*}{LSTUR}      & No Defense               & 66.67±0.09 & 32.30±0.12 & 34.97±0.11 & 40.67±0.11 & 65.17±0.04 & 31.91±0.08 & 34.39±0.13 & 42.19±0.08 \\
                            & Median~\cite{yin2018byzantine}                   & 56.26±0.18 & 25.65±0.19 & 26.77±0.19 & 32.35±0.16 & 60.22±0.12 & 27.93±0.13 & 29.35±0.13 & 37.45±0.12  \\
                            & Trimmed-Mean~\cite{yin2018byzantine}             & 63.19±0.10 & 29.58±0.07 & 31.66±0.07 & 37.41±0.07 & 61.48±0.29 & 29.02±0.06 & 30.69±0.06 & 38.68±0.12  \\
                            & Krum~\cite{NIPS2017_f4b9ec30}                    & 56.62±0.41 & 25.69±0.48 & 26.97±0.59 & 32.55±0.54 & 62.71±0.16 & 29.95±0.20 & 31.99±0.19 & 39.97±0.13  \\
                            & Multi-Krum~\cite{NIPS2017_f4b9ec30}             & 65.94±0.19 & 31.68±0.15 & 34.19±0.15 & 39.92±0.14 & 62.86±0.11 & 29.97±0.08 & 31.90±0.09 & 39.96±0.09  \\
                            & Norm-Bounding~\cite{sun2019can}           & 66.75±0.16 & 32.30±0.18 & 34.96±0.20 & 40.66±0.18 & 65.22±0.14 & 31.98±0.09 & 34.49±0.09 & 42.27±0.10  \\ \Xhline{1.0pt}                 
\end{tabular}
}
\label{tab:exp-defenses}
\end{table*}

\subsection{Dataset and Experimental Settings}
\label{sec:dataset}
We conduct experiments on two real-world datasets: MIND\footnote{\url{https://msnews.github.io/}. We use the small version of MIND for fast experiments.} and Feeds.
MIND is a public dataset collected from anonymized behavior logs of Microsoft news website, which contains user behaviors in six weeks.
We collect the Feeds dataset from Microsoft news App from August 1st, 2020 to September 1st, 2020.
For MIND, we directly use the provided training, validation, and test datasets.
For Feeds, we use the impressions in the first three weeks as the training dataset, the impressions in the later two days as the validation dataset, and the rest in the last week for testing.
The detailed dataset statistics are summarized in Table~\ref{tab:stat}.
Following previous news recommendation works ~\cite{qi-etal-2021-pp,wu-etal-2019-neural-news,an-etal-2019-neural}, we use AUC, MRR, nDCG@5 and nDCG@10 as the evaluation metrics. We note that the experimental results reported here are all on benign datasets.

We evaluate our UA-FedRec against two news recommendation models: NRMS~\cite{wu-etal-2019-neural-news} and LSTUR~\cite{an-etal-2019-neural}.
We apply the non-uniform weighted averaging FedAdam~\cite{reddi2021adaptive} to train the news recommendation models.
We set $\lambda_1$, $\lambda_2$ and $\lambda_3$ to 1.5, 1.5, 3, respectively, on Feeds with LSTUR.
In other experiments, $\lambda_1$, $\lambda_2$ and $\lambda_3$ are set to 3.0.
The dimension of news representations is 400.
We apply dropout with dropout rate 0.2 to mitigate overfitting,
The learning rate is 0.0001.
The number of negative samples associated with each positive sample is 4.
The number of clients randomly sampled per round is 50 for both MIND and Feeds.
The percentage of malicious clients is set to 1\%.
All hyper-parameters are selected according to results on the validation set.
We repeat each experiment 5 times independently, and report the average results with standard deviations.

\subsection{Performance Comparison}

We select existing untargeted attacks as baseline methods and compare our UA-FedRec with them. The baseline methods include the following data poisoning attack methods:
\begin{itemize}
\item \textbf{No Attack}, where no attack is applied. It is the upper bound of model performance;
\item \textbf{Label Flipping (LF)}~\cite{10.5555/3489212.3489304}, flipping click labels of training input samples;
\item \textbf{Popularity Perturbation (Pop)}~\cite{zhang2021pipattack}, an untargeted version of the explicit boosting in PipAttack, where malicious clients click only cold news;
\item \textbf{FedAttack}~\cite{10.1145/3534678.3539119}, sampling items with representations that closely resemble the representations of the malicious user as negative samples.

\end{itemize}
and the following model poisoning attack methods:
\begin{itemize}
\item \textbf{Gaussian}~\cite{10.5555/3489212.3489304}, sampling malicious updates from a Gaussian distribution estimated from benign model updates;
\item \textbf{Little is Enough (LIE)}~\cite{NEURIPS2019_ec1c5914}, adding a small amount of noise to each dimension of the average of the benign updates;
\item \textbf{Fang}~\cite{10.5555/3489212.3489304}, where noise is added in the opposite direction from the average of benign model updates.
\end{itemize}

\begin{table*}[!t]
\centering
\caption{Attack performance of different methods with the Norm-Bounding defense.}
\scalebox{0.93}{
\begin{tabular}{c|c|cccc|cccc}
\Xhline{1.0pt}
 Base                       & \multirow{2}{*}{Methods} & \multicolumn{4}{c|}{MIND}                         & \multicolumn{4}{c}{Feeds}    \\ \cline{3-10} 
 Model                      &                          & AUC        & MRR        & nDCG@5     & nDCG@10    & AUC & MRR & nDCG@5 & nDCG@10 \\ \hline
\multirow{6}{*}{NRMS}       & No Attack         & 66.92±0.19 & 32.44±0.13 & 35.18±0.14 & 40.88±0.14 & 64.97±0.05 & 31.84±0.09 & 34.31±0.10 &                                   42.08±0.09 \\
                            & LF~\cite{10.5555/3489212.3489304}           & 66.70±0.11 & 32.27±0.12 & 34.99±0.14 & 40.69±0.12 & 65.02±0.10 & 31.87±0.13 & 34.34±0.15 & 42.11±0.13 \\
                            & Pop~\cite{zhang2021pipattack}          & 66.68±0.25 & 32.30±0.16 & 35.00±0.19 & 40.68±0.18 & 65.15±0.10 & 31.97±0.12 & 34.43±0.12 &       42.22±0.11 \\
                            & FedAttack~\cite{10.1145/3534678.3539119} & 66.73±0.01 & 32.36±0.02 & 35.09±0.04 & 40.77±0.04 & 65.15±0.13 & 31.93±0.08 & 34.41±0.13 & 42.17±0.10 \\
                            & Gaussian~\cite{10.5555/3489212.3489304}     & 66.66±0.07 & 32.28±0.08 & 34.98±0.10 & 40.69±0.09 & 65.02±0.06 & 31.90±0.04 & 34.34±0.08 &       42.13±0.04 \\
                            & LIE~\cite{NEURIPS2019_ec1c5914}          & 63.46±0.26 & 30.07±0.19 & 32.32±0.23 & 38.03±0.22 & 61.21±0.34 & 28.83±0.25 & 30.47±0.30 & 38.46±0.31 \\
                            & Fang~\cite{10.5555/3489212.3489304}         & 66.25±0.19 & 32.03±0.18 & 34.65±0.22 & 40.35±0.21 & 63.08±0.70 & 30.26±0.59 & 32.24±0.76 & 40.16±0.69 \\ \cline{2-10} 
                            & UA-FedRec & \textbf{57.00}±0.26 & \textbf{25.61}±0.21 & \textbf{26.91}±0.24 & \textbf{32.57}±0.25 & \textbf{59.73}±0.26 & \textbf{27.62}±0.24 & \textbf{28.98}±0.30 & \textbf{37.08}±0.27 \\ \hline
\multirow{6}{*}{LSTUR}      & No Attack         & 66.75±0.16 & 32.30±0.18 & 34.96±0.20 & 40.66±0.18 & 65.22±0.14 & 31.98±0.09 & 34.49±0.09 & 42.27±0.10 \\
                            & LF~\cite{10.5555/3489212.3489304} & 66.62±0.14 & 32.24±0.11 & 34.90±0.11 & 40.61±0.10 & 65.02±0.07 & 31.85±0.07 & 34.29±0.08 & 42.11±0.05 \\
                            & Pop~\cite{zhang2021pipattack}          & 66.75±0.09 & 35.04±0.09 & 35.05±0.12 & 40.74±0.10 & 65.25±0.06 & 32.02±0.02 & 34.52±0.03 & 42.31±0.04 \\
                            & FedAttack~\cite{10.1145/3534678.3539119} & 66.78±0.04 & 32.42±0.06 & 35.09±0.07 & 40.77±0.07 & 65.18±0.16 & 31.89±0.11 & 34.37±0.15 & 42.18±0.12 \\
                            & Gaussian~\cite{10.5555/3489212.3489304}    & 66.69±0.25 & 32.29±0.18 & 34.94±0.21 & 40.64±0.20 & 65.18±0.05 & 31.93±0.05 & 34.39±0.05 & 42.19±0.06 \\
                            & LIE~\cite{NEURIPS2019_ec1c5914}      & 64.97±0.12 & 31.11±0.04 & 33.60±0.03 & 39.30±0.05 & 64.95±0.28 & 31.78±0.25 & 34.23±0.26 & 42.06±0.21 \\
                            & Fang~\cite{10.5555/3489212.3489304}         & 66.36±0.12 & 32.09±0.10 & 34.70±0.09 & 40.40±0.09 & 64.83±0.13 & 31.63±0.15 & 34.04±0.17 & 41.86±0.15 \\ \cline{2-10} 
                            & UA-FedRec & \textbf{55.24}±0.85 & \textbf{24.89}±0.51 & \textbf{25.99}±0.54 & \textbf{31.56}±0.56  & \textbf{61.83}±0.87 & \textbf{29.10}±0.78 & \textbf{30.90}±0.95 & \textbf{38.92}±0.90  \\ \Xhline{1.0pt}
\end{tabular}
}
\label{tab:exp-norm-bound}
\end{table*}
\begin{table*}[!t]
\centering
\caption{Attack performance of different methods with the Multi-Krum defense.}
\scalebox{0.93}{
\begin{tabular}{c|c|cccc|cccc}
\Xhline{1.0pt}
    Base                    & \multirow{2}{*}{Methods} & \multicolumn{4}{c|}{MIND}                         & \multicolumn{4}{c}{Feeds}    \\ \cline{3-10} 
   Model                    &                          & AUC        & MRR        & nDCG@5     & nDCG@10    & AUC & MRR & nDCG@5 & nDCG@10 \\ \hline
\multirow{6}{*}{NRMS}       & No Attack         & 65.80±0.17 & 31.66±0.10 & 34.23±0.11 & 39.93±0.12 & 62.51±0.08 & 29.73±0.06 & 31.62±0.07 &                                   39.63±0.08 \\
                            & LF~\cite{10.5555/3489212.3489304}           & 65.63±0.28 & 31.54±0.21 & 34.06±0.24 & 39.77±0.22 & 62.44±0.07 & 29.74±0.07 & 31.64±0.07 & 39.64±0.07 \\
                            & Pop~\cite{zhang2021pipattack}          & 65.73±0.19 & 31.62±0.16 & 34.14±0.23 & 39.83±0.21 & 62.28±0.11 & 29.57±0.13 & 31.39±0.17 & 39.44±0.14 \\
                            & FedAttack~\cite{10.1145/3534678.3539119} & 65.85±0.17 & 31.67±0.11 & 34.20±0.14 & 39.90±0.14 & 62.51±0.03 & 29.76±0.04 & 31.62±0.05 & 39.64±0.05 \\
                            & Gaussian~\cite{10.5555/3489212.3489304}     & 65.75±0.18 & 31.66±0.19 & 34.19±0.17 & 39.90±0.16 & 62.28±0.16 & 29.59±0.14 & 31.42±0.19 & 39.46±0.13 \\
                            & LIE~\cite{NEURIPS2019_ec1c5914}   & 62.93±0.23 & 29.64±0.09 & 31.77±0.10 & 37.49±0.10 & 61.77±0.06 & 29.20±0.06 & 30.89±0.07 & 38.96±0.06 \\
                            & Fang~\cite{10.5555/3489212.3489304}         & 65.45±0.16 & 31.37±0.12 & 33.83±0.13 & 39.53±0.15 & 61.84±0.28 & 29.26±0.28 & 30.99±0.36 & 39.02±0.29 \\ \cline{2-10} 
                            & UA-FedRec       & \textbf{60.30}±0.80 & \textbf{27.97}±0.47 & \textbf{29.78}±0.51 & \textbf{35.39}±0.52 & \textbf{61.02}±0.10    & \textbf{28.65}±0.11 & \textbf{30.20}±0.15 & \textbf{38.26}±0.13        \\ \hline
\multirow{6}{*}{LSTUR}      & No Attack         & 65.94±0.19 & 31.68±0.15 & 34.19±0.15 & 39.92±0.14 & 62.86±0.11 & 29.97±0.08 & 31.90±0.09 & 39.96±0.09 \\
                            & LF~\cite{10.5555/3489212.3489304}           & 66.06±0.08 & 31.82±0.07 & 34.33±0.08 & 40.06±0.08 & 62.54±0.07 & 29.74±0.07 & 31.62±0.10 & 39.69±0.08 \\
                            & Pop~\cite{zhang2021pipattack}          & 65.97±0.18 & 31.79±0.15 & 34.30±0.17 & 40.02±0.16 & 62.40±0.07 & 29.55±0.23 & 31.37±0.29 & 39.49±0.30       \\
                            & FedAttack~\cite{10.1145/3534678.3539119} & 65.92±0.14 & 31.79±0.13 & 34.32±0.13 & 40.03±0.13 & 62.81±0.22 & 30.01±0.16 & 31.95±0.20 & 39.94±0.20 \\
                            & Gaussian~\cite{10.5555/3489212.3489304}     & 65.99±0.18 & 31.76±0.13 & 34.26±0.15 & 39.99±0.13 & 62.81±0.09 & 29.91±0.01 & 31.84±0.04 & 39.90±0.02 \\
                            & LIE~\cite{NEURIPS2019_ec1c5914}          & 65.92±0.18 & 31.23±0.17 & 33.60±0.17 & 39.45±0.12 & 62.51±0.16 & 29.78±0.17 & 31.70±0.17 & 39.71±0.21 \\
                            & Fang~\cite{10.5555/3489212.3489304}      & 65.69±0.26 & 31.58±0.16 & 34.06±0.17 & 39.78±0.18 & 62.11±0.01 & 29.40±0.02 & 31.19±0.01 & 39.27±0.04 \\ \cline{2-10} 
                            & UA-FedRec & \textbf{59.87}±0.62 & \textbf{27.24}±0.31 & \textbf{28.89}±0.32 & \textbf{34.63}±0.32 & \textbf{61.70}±0.31 &  \textbf{29.19}±0.03 & \textbf{30.92}±0.09 &  \textbf{38.93}±0.17   \\ \Xhline{1.0pt}
\end{tabular}
}
\label{tab:exp-multikrum}
\end{table*}
The experimental results are shown in Table~\ref{tab:exp-no-defense}.
We have the following observations from the table.
First, our UA-FedRec outperforms the data poisoning attack methods (LF, Pop, FedAttack).
This is because UA-FedRec directly manipulates model updates, while data poisoning attacks perturb only input samples.
Second, our UA-FedRec outperforms other model poisoning attack methods (Gaussian, LIE, and Fang).
This is because UA-FedRec has fully exploited the prior knowledge in news recommendation and federated learning: it applies both user model perturbation and news similarity perturbation since
user modeling and news modeling are critical for news recommendation.
The user model perturbation makes updates of user model less accurate.
The news similarity perturbation makes similar news farther and dissimilar news closer, which can effectively interfere with news modeling.
Moreover, UA-FedRec applies quantity perturbation to amplify the impact of malicious updates.
Third, the well-designed model poisoning attacks (LIE, Fang, and UA-FedRec) perform better than the data poisoning attacks (LF and Pop).
This is because perturbing model updates is more effective than manipulating input samples.
A model poisoning attack is generally more flexible and performs better than a data poisoning attack.
Finally, our UA-FedRec significantly degrades the performance of news recommendation models with only 1\% of malicious clients, making the attack more practical for the federated news recommendation.

\begin{figure*}[!t]
  \centering
  \includegraphics[width=0.99\textwidth]{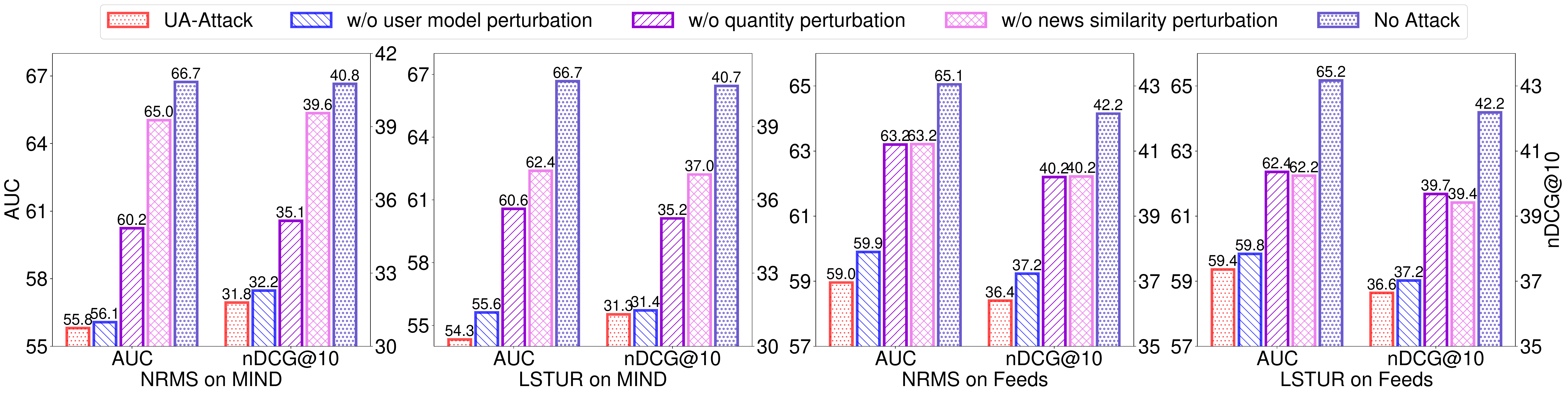}
  \caption{Impact of core components in UA-FedRec. }
  \label{fig:ablation}
\end{figure*}
\subsection{Circumventing Defenses}
To evaluate the effectiveness of existing defenses against our UA-FedRec, we consider several popular defenses, including:
\begin{itemize}
\item \textbf{Median}~\cite{yin2018byzantine}, a coordinate-wise defense that aggregates updates by computing the median of each dimension of the updates.
\item \textbf{Trimmed-Mean}~\cite{yin2018byzantine}, a coordinate-wise defense that aggregates updates by computing the trimmed mean of each dimension of the updates.
\item \textbf{Krum}~\cite{NIPS2017_f4b9ec30}, selecting the update from the set of received updates that is closest to its subset of neighboring updates.
\item \textbf{Multi-Krum}~\cite{NIPS2017_f4b9ec30}, selecting multiple neighbors with the scores of Krum, and averaging the selected updates.
\item \textbf{Norm-Bounding}~\cite{sun2019can}, truncating the norm of received updates and applying non-uniform aggregation.
\end{itemize}

A defense method should not incur any significant adverse impact on the performance of a model. To evaluate the impact of these defenses on the performance of federated news recommendation systems, we first evaluate them with NRMS and LSTUR news recommendation models on both datasets. The experimental results are shown in Table~\ref{tab:exp-defenses}. The table shows that some defenses (Krum, Median, Trimmed-Mean) severely degrade the performance of both news recommendation models. As a result, we select only the defenses, i.e., Norm-Bounding and Multi-Krum, that have small performance degradation to evaluate our UA-FedRec and the baseline methods.

The experimental results of attacking federated new recommendation systems are shown in Table~\ref{tab:exp-norm-bound} when the Norm-Bounding defense is applied and in Table~\ref{tab:exp-multikrum} when the Multi-Krum defense is applied.
From both Table~\ref{tab:exp-norm-bound} and Table~\ref{tab:exp-multikrum}, we have several observations.
First, 
data poisoning attacks (LF, Pop and FedAttack) are ineffective when Norm-Bounding or Multi-Krum is applied. These attacks perform poorly without any defense, as Table~\ref{tab:exp-no-defense} shows, since they require more than 1\% malicious clients, let alone with defense.
Second, our UA-FedRec outperforms the model poisoning attacks (LIE and Fang) with both Norm-Bounding and Multi-Krum defenses. Our news similarity perturbation and user model perturbation can still effectively impact news recommendation models even when these defenses are applied.  
Third, well-designed model poisoning attacks (LIE, Fang, and UA-FedRec) perform better than data poisoning attacks (LF and Pop).
This is because these model poisoning attack methods optimize the perturbation degree directly on model updates while adding constraints to circumvent defenses, resulting in a better capability to evade defenses.
Forth, compared with the performance without any defense, both Norm-Bounding and Multi-Krum improve the performance when facing the tested attacks, except for Multi-Krum on Feeds.
This is because the defenses can mitigate the impact of malicious updates or directly detect malicious updates and filter them out.


Our experimental results indicate that existing robust aggregation rules either significantly degrade the performance of news recommendation models or cannot effectively thwart UA-FedRec. As a future work, we plan to study effective defense methods on federated news recommendation systems to defend against UA-FedRec. Specifically, first, we plan to detect malicious news similarity updates to defend against the news similarity perturbation.
Since the news information is public for both server and clients, the server can estimate news similarity scores with self-supervised or unsupervised training methods.
Second, we plan to take sample sizes into robust aggregation rules to restrict the impact of updates with larger sample sizes to defend against quantity perturbation.
Third, we plan to detect malicious user modeling updates to defend against user perturbation.

\subsection{Ablation Study}
\label{sec:ablation}

\begin{figure}
     \centering
     \begin{subfigure}[b]{0.232\textwidth}
         \centering
         \includegraphics[width=\textwidth]{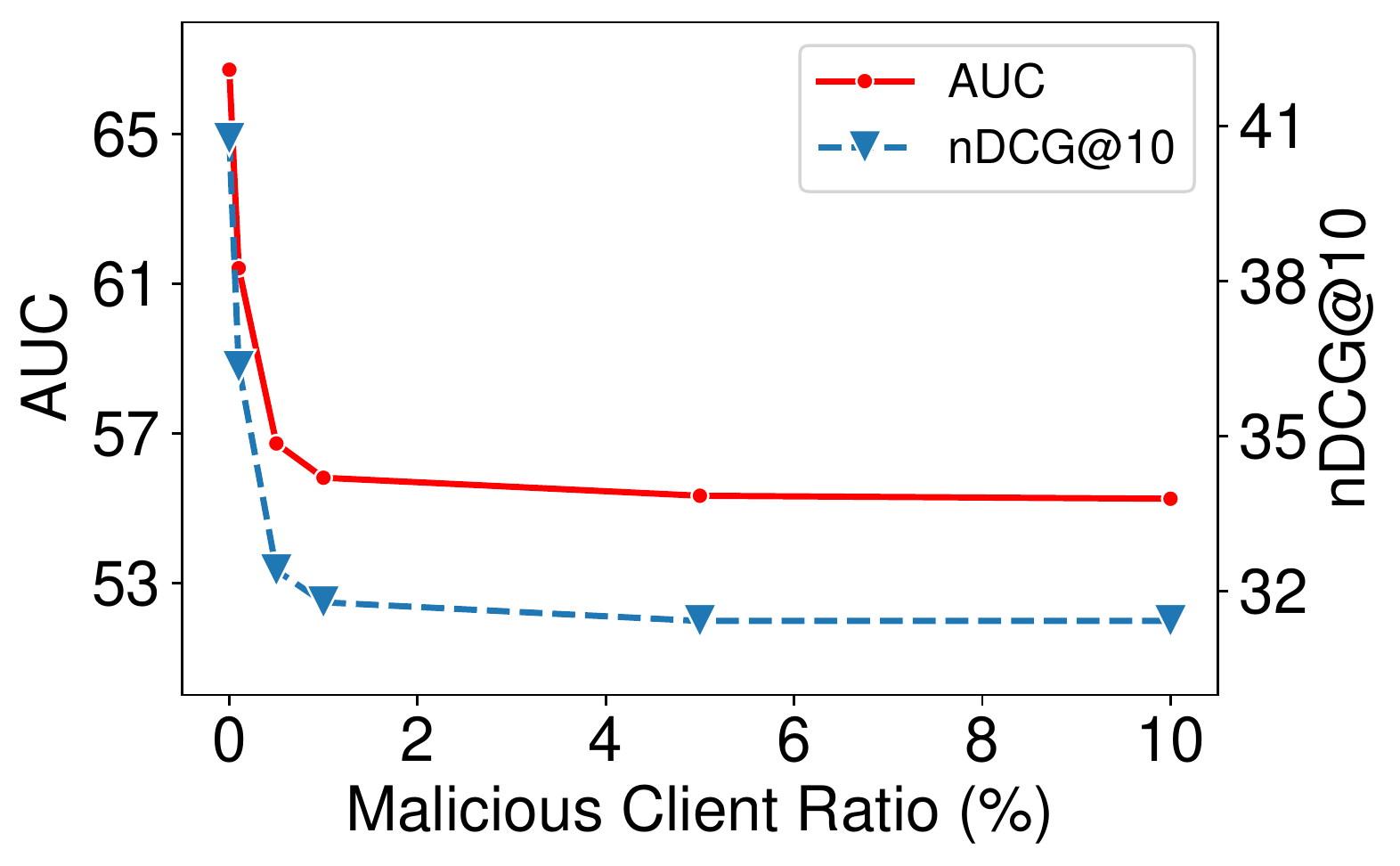}
         \caption{NRMS on MIND.}
         \label{fig:mal_ratio_nrms}
     \end{subfigure}
     \hfill
     \begin{subfigure}[b]{0.232\textwidth}
         \centering
         \includegraphics[width=\textwidth]{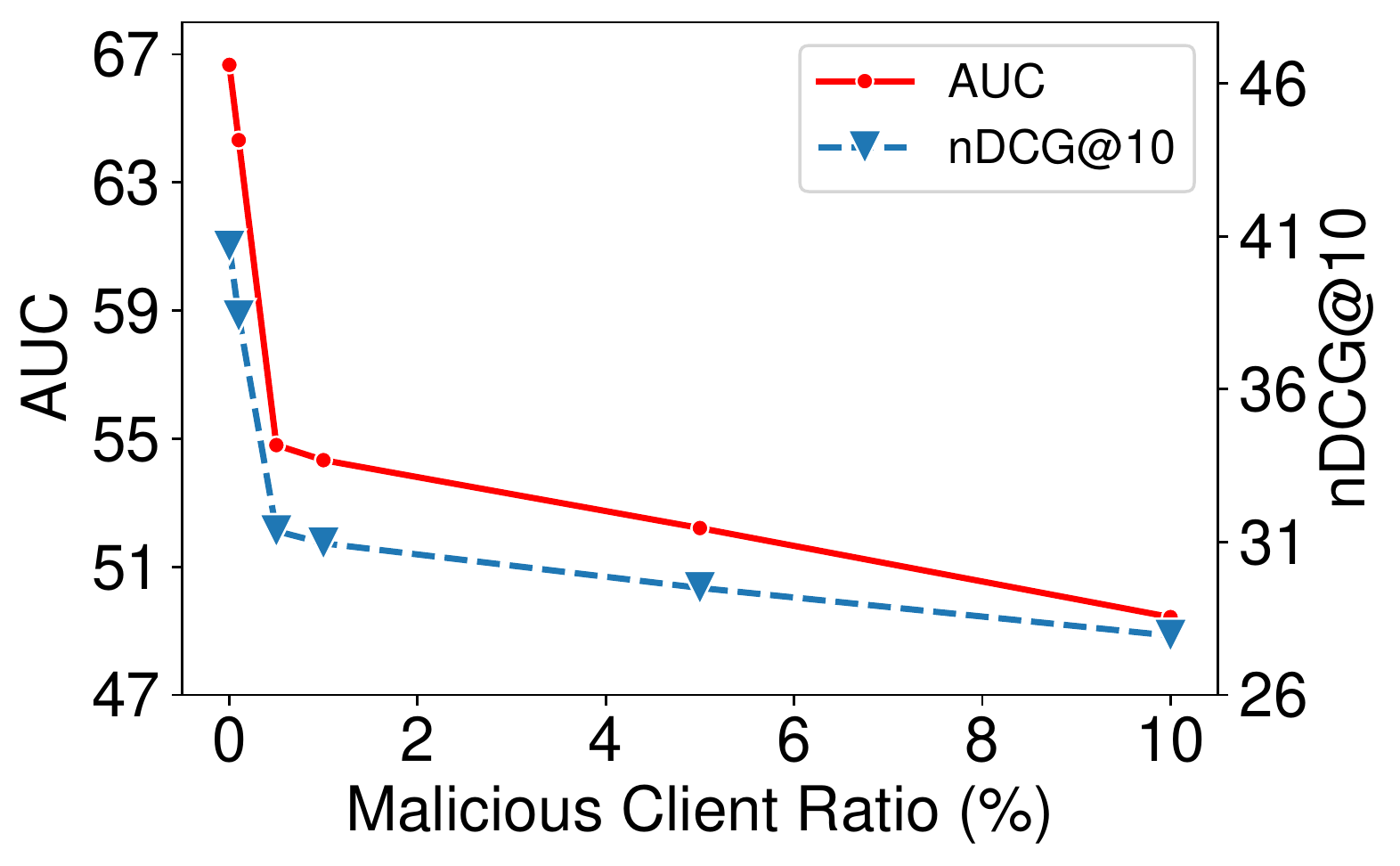}
         \caption{LSTUR on MIND.}
         \label{fig:mal_ratio_lstur}
     \end{subfigure}
     \centering
     \begin{subfigure}[h]{0.235\textwidth}
         \centering
         \includegraphics[width=\textwidth]{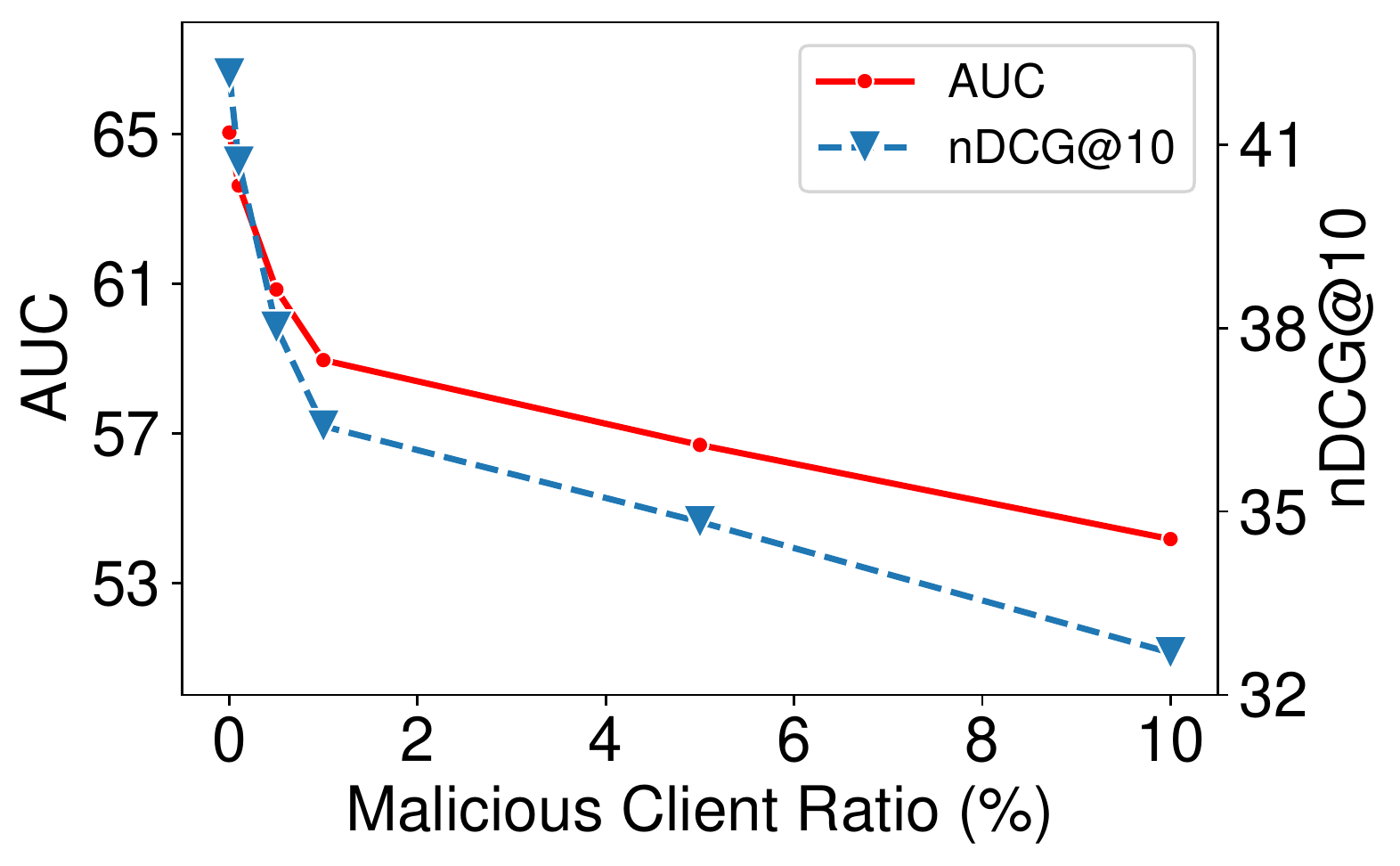}
         \caption{NRMS on Feeds.}
         \label{fig:mal_ratio_nrms_feeds}
     \end{subfigure}
     \hfill
     \begin{subfigure}[h]{0.235\textwidth}
         \centering
         \includegraphics[width=\textwidth]{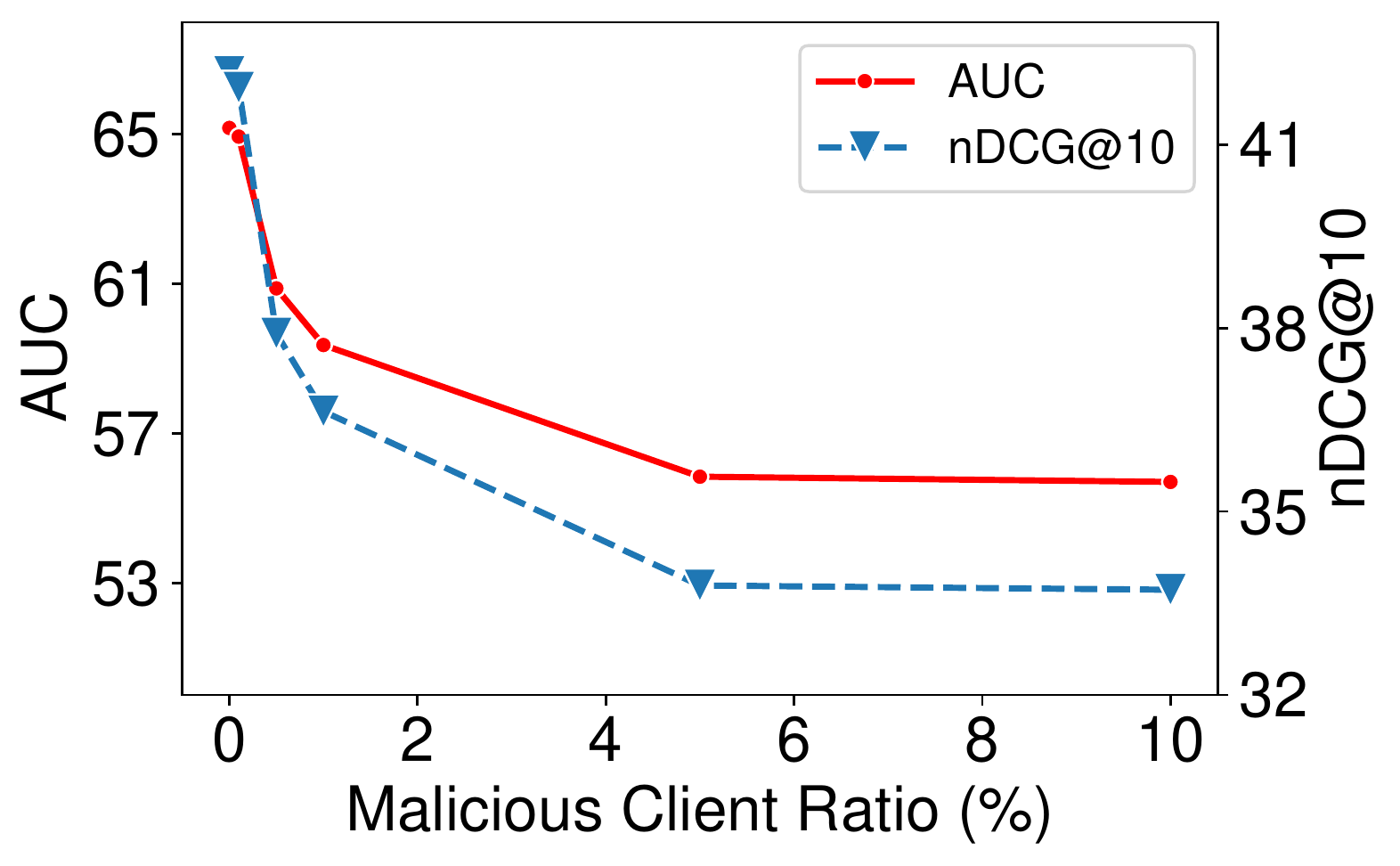}
         \caption{LSTUR on Feeds.}
         \label{fig:mal_ratio_lstur_feeds}
     \end{subfigure}
    \caption{Impact of malicious client ratio on MIND.}
    \label{fig:mal_ratio}
\end{figure}
In this subsection, we study the impact of the three core components of our UA-FedRec, i.e., user model perturbation, news similarity perturbation, and quantity perturbation.
The experimental results on MIND and Feeds are shown in Figure~\ref{fig:ablation}.
We can make the following observations.
First, the attack performance of our UA-FedRec degrades without the news similarity perturbation.
This is because news similarity modeling is critical to news recommendation and our news similarity perturbation can effectively interfere with model's learning news similarity.
Second, the attack performance of our UA-FedRec degrades without the quantity perturbation.
This is because model updates are aggregated based on sample sizes in FedAvg.
Our quantity perturbation amplifies the impact of malicious updates.
Third, the attack performance of our UA-FedRec degrades a little without the user perturbation.
Our user perturbation manipulates a user model update in the opposite direction of the average of benign updates. Since news representations are polluted by the news similarity perturbation, the user model is unable to capture user interests even without the user model perturbation, resulting in a small drop of performance without the user perturbation.

\subsection{Impact of Malicious Client Ratio}
\label{sec:mal_ratio}
In this subsection, we study the impact of the percentage of malicious clients.
We conduct experiments with 0.1\%, 0.5\%, 1\%, 5\% and 10\% of malicious clients.
The experimental results on MIND and Feeds are shown in Figure~\ref{fig:mal_ratio}.
We can see that the attack performance improves with a larger percentage of malicious clients.
This is in line with expectations since more malicious updates are uploaded with a higher percentage of malicious clients, resulting in a more heavily affected global news recommendation model.
Second, our UA-FedRec can effectively attack the global news recommendation model with a percentage of malicious clients as low as 0.1\%. By exploiting the prior knowledge in news recommendation and federated learning, UA-FedRec effectively perturbs news similarity modeling and user modeling and amplifies the impact of malicious updates with quantity perturbation.
These perturbations can effectively reduce the percentage of malicious clients launching an effective untargeted attack. 

\subsection{Impact of Known News}
In this subsection, we study how the number of pieces of news known to the attacker affects the attack performance.
We conduct some experiments on MIND with NRMS.
The experimental results are shown in Table~\ref{tab:news-ratio}.
We have several observations from the results.
First, our UA-FedRec has great attack performance even with a subset of news known to the attacker, which shows it is enough for the attacker to perturb news similarity modeling with the partial knowledge of news items.
Second, the attack performance drops a little with lower ratios of news known to the attacker.
This is because, with more pieces of news, the attacker can find dissimilar news and similar news more accurately.

\begin{table}[!t]
\caption{Performance of NRMS with different ratios of news known to the attacker on MIND.}
\scalebox{1.0}{
\begin{tabular}{c|cccc}
\Xhline{1.0pt}
Ratio & AUC   & MRR   & nDCG@5 & nDCG@10 \\ \hline
0.5   & 54.70 & 24.47 & 25.40  & 31.05   \\
0.2   & 55.36 & 25.26 & 26.24  & 31.96   \\
0.1   & 56.02 & 25.57 & 26.79  & 32.32   \\ \Xhline{1.0pt}
\end{tabular}
}
\label{tab:news-ratio}
\end{table}

\subsection{Hyper-parameter Analysis}

\begin{figure}[!t]
  \centering
  \includegraphics[width=0.33\textwidth]{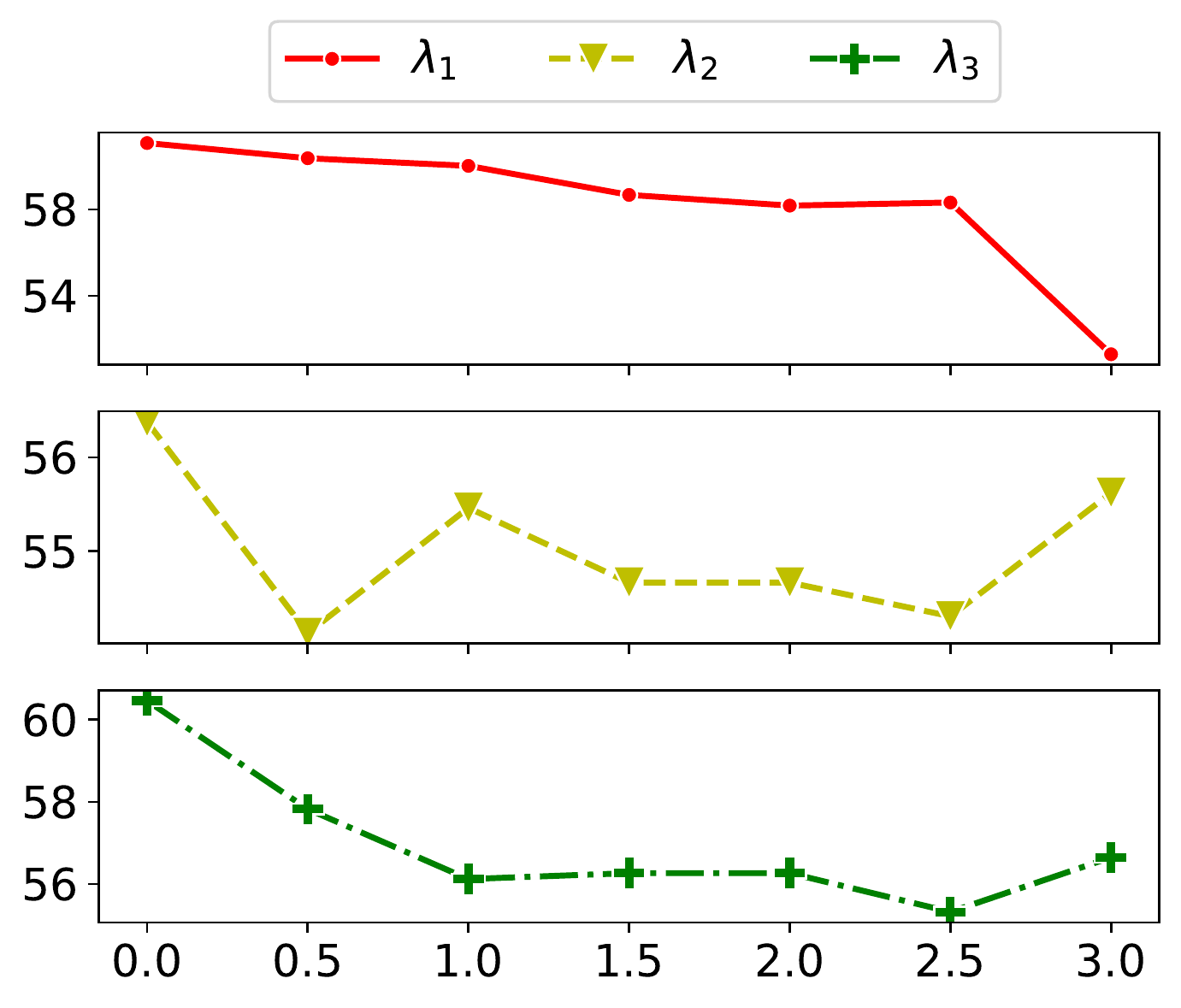}
  \caption{Impact of hyper-parameters in UA-FedRec on NRMS on MIND with no defense. }
  \label{fig:hyper}
\end{figure}

In this subsection, we analyze the impact of three important hyper-parameters in UA-FedRec, i.e., $\lambda_1$, $\lambda_2$ and $\lambda_3$.
The empirical analysis of the impact of hyper-parameters on NRMS on MIND with no defense is shown in Figure~\ref{fig:hyper}, where we can have several observations.
First, the attack performance without defense improves with higher $\lambda_1$.
This is because the norm of malicious news encoder updates increases with larger $\lambda_1$, thereby having a more significant impact on model performance.
Second, $\lambda_2$ has a relatively minimal impact on attack performance. This is because, given the news representations' pollution due to news similarity perturbation, the user model fails to capture user interests, even without user model perturbation.
Third, the attack performance under no defense improves with larger $\lambda_3$.
This is because with larger $\lambda_3$, the weight of malicious updates increases.
Finally, since larger $\lambda_1$, $\lambda_2$ and $\lambda_3$ would make the updates and sample size of the malicious users more different from those of the benign users, we choose the hyper-parameters based on the trade-off between the attack performance and covertness.
The detailed settings of hyper-parameters are shown in Appendix~\ref{sec:hyper}.

\section{Conclusion}
In this paper, we propose an untargeted attack, called UA-FedRec, on federated news recommendation systems.
By exploiting the prior knowledge in news recommendation and federated learning, we have designed three perturbation methods in UA-FedRec, i.e., news similarity perturbation, user model perturbation and quantity perturbation, to interfere with news similarity modeling, user modeling, and amplify the impact of malicious updates, respectively. The user model perturbation makes news representations of similar news farther and those of dissimilar news closer, which can effectively interfere with news similarity modeling in news recommendation.
The user model perturbation perturbs user model updates in opposite directions of benign updates to interfere with user modeling.  
The quantity perturbation enlarges sample sizes of malicious clients in a reasonable range to amplify the impact of malicious updates. 
Extensive experiments on two real-world datasets indicate that our UA-FedRec can effectively degrade the performance of federated news recommendation systems while circumventing defenses with a percentage of malicious clients as low as 1\%. It outperforms existing untargeted attacks using data poisoning or model poisoning. Our study reveals a critical security issue in existing federated news recommendation systems and calls for more research efforts to address this issue. 
In the future, we plan to study effective defense methods to thwart UA-FedRec and other potential attacks against news recommendation systems.
In addition, we also plan to extend our UA-FedRec to other content-based recommendation scenarios.
\section*{Acknowledgements}
We would like to thank Hao Wang, Yang Yu and Chao Zhang for their great comments on paper writing.

\bibliographystyle{ACM-Reference-Format}
\bibliography{main}

\appendix

\section*{Appendix}

\subsection*{Stable Vector Distances}
\label{sec:distances-no-change}
We conduct an empirical study to evaluate whether the news similarity does not change significantly in $K=100$ rounds.
We first randomly sample 10k news pairs from MIND and compute the similarity scores with two NRMS models saved at $M$ and $M+100$ rounds, respectively. Then we compute the Pearson correlation coefficient of the similarity scores from the two models.
Our results in Table~\ref{tab:stable-news-sim} show a very strong positive linear relationship between the two similarity scores, indicating that the news similarity does not change significantly in $K=100$ rounds.

\begin{table}[H]
\centering
\caption{Pearson correlation coefficient of the similarity scores from 10k news pairs on MIND with the NRMS models saved at $M$ and $M+100$ rounds.}
\scalebox{0.85}{
\begin{tabular}{c|ccccc}
\Xhline{1.5pt}
 M & 1000	& 2000	& 3000	& 4000	& 5000 \\
 r	& 0.9709	& 0.9749	& 0.9869	& 0.9908	& 0.9870 \\
\Xhline{1.5pt}
\end{tabular}
}
\label{tab:stable-news-sim}
\end{table}

\subsection*{Hyper-parameter Settings}
\label{sec:hyper}
The complete hyper-parameter settings on MIND are listed in Table~\ref{tab:hyper-mind}, and the complete hyper-parameter settings on Feeds are listed in Table~\ref{tab:hyper-feeds}.
\begin{table}[H]
\centering
\caption{Hyper-parameter settings on MIND.}
\scalebox{0.85}{
\begin{tabular}{c|c|c}
\Xhline{1.5pt}
Hyperparameters             & NRMS       & LSTUR  \\ \hline
learning rate               & 0.0001     & 0.0001 \\
number of negative samples $P$       & 4          & 4      \\
sampled user per round $k$  & 50         & 50  \\
number of rounds to update news neighbors $K$ & 100 & 100 \\
malicious clients number $m$ & 500       & 500    \\
dimention of news representations  & 400    & 400    \\
dropout rate            & 0.2      & 0.2   \\
$\lambda_1$                & 3.0      & 1.5   \\
$\lambda_2$                & 3.0      & 1.5   \\
$\lambda_3$                & 3.0      & 3.0   \\ 
Adam $\beta_1$            & 0.9        & 0.9    \\
Adam $\beta_2$             & 0.99       & 0.99   \\
Adam $\tau$               & $10^{-8}$  &$10^{-8}$\\
\Xhline{1.5pt}
\end{tabular}
}
\label{tab:hyper-mind}
\end{table}

\begin{table}[H]
\centering
\caption{Hyper-parameter settings on Feeds.}
\scalebox{0.85}{
\begin{tabular}{c|c|c}
\Xhline{1.5pt}
Hyperparameters             & NRMS       & LSTUR  \\ \hline
learning rate               & 0.0001     & 0.0001 \\
number of negative samples $P$       & 4          & 4      \\
sampled user per round $k$  & 50         & 50  \\
number of rounds to update news neighbors $K$ & 100 & 100 \\
malicious clients number $m$ & 100       & 100    \\
dimention of news representations  & 400    & 400    \\
dropout rate            & 0.2      & 0.2   \\
$\lambda_1$                & 3.0      & 3.0   \\
$\lambda_2$                & 3.0      & 3.0   \\
$\lambda_3$                & 3.0      & 3.0   \\ 
Adam $\beta_1$            & 0.9        & 0.9    \\
Adam $\beta_2$             & 0.99       & 0.99   \\
Adam $\tau$               & $10^{-8}$  &$10^{-8}$\\
\Xhline{1.5pt}
\end{tabular}
}
\label{tab:hyper-feeds}
\end{table}

\subsection*{Experimental Environment}

There are 8 Tesla V100-SXM2-32GB in the server with CUDA 11.1.
The CPU is Intel(R) Xeon(R) Platinum 8168 CPU @ 2.70GHz.
We use python 3.7.11, pytorch 1.10.0.
Each experiment is run on a single GPU and a single CPU core.

\end{document}